\documentclass[12pt, a4paper]{amsart}
\usepackage{amsmath,amssymb}
\usepackage[dvipdfmx]{graphicx}
\usepackage{wrapfig}
\newtheorem{theo}{Theorem}[section]
\newtheorem{prop}{Proposition}[section]
\newtheorem{defi}{Definition}[section]
\newtheorem{lemm}{Lemma}[section]
\newtheorem{exam}{Example}[section]
\newtheorem{coro}{Corollary}[section]
\newtheorem{rema}{Remark}[section]
\textwidth=\paperwidth
\advance\textwidth by -5cm
\oddsidemargin=\paperwidth
\advance\oddsidemargin by -\textwidth
\divide\oddsidemargin by 2
\advance\oddsidemargin by -1in

\textheight=\paperheight
\advance\textheight by -5.8cm
\advance\topmargin by -1in
\advance\topmargin by -\headheight
\advance\topmargin by -\headsep
\makeatletter

\@addtoreset{equation}{section}
\renewcommand\theequation{\@arabic\c@section.\@arabic\c@equation}
\makeatother

\begin{document}
\title[Periodic udKdV equation and background solutions]{The generalized periodic ultradiscrete KdV equation and its background solutions}
\author{Masataka Kanki}
\address{Graduate school of Mathematical Sciences, University of Tokyo, 3-8-1 Komaba, Tokyo 153-8914, Japan}
\email{kanki@ms.u-tokyo.ac.jp}
\begin{abstract}
We investigate the ultradiscrete KdV equation with periodic boundary conditions where the two parameters (capacity of the boxes and that of the carrier) are arbitrary integers. We give a criterion to allow a periodic boundary condition when initial states take arbitrary integer values. Conserved quantities are constructed for the periodic systems. Construction of background solutions of the periodic ultradiscrete KdV equation from the Jacobi theta function is also presented.
\end{abstract}
\maketitle
\section{Preface}
In this paper we investigate the boundary conditions and the background solutions of a cellular automaton which is called the Box Ball System (BBS). 

The content of this paper is as follows. The BBS is derived from the discrete KdV equation by a limiting procedure called `ultradiscretization', which we will explain briefly in section 2.
In section 3, we give a criterion for the BBS and the BBS with a Carrier (BBSC) to allow a periodic boundary condition even when the system allows `negative solitons'. 
We also define the background solution of the BBSC using the conserved quantities of the BBSC. In section 4, we introduce a gauge transformed discrete KdV equation whose solutions converge to the upward-shifted solutions of the BBSC by ultradiscretization. 
We then investigate the relation between the Jacobi theta solutions of the discrete KdV equation and the background solutions of the BBS in section 5. In section 6, we present an example of multi-soliton solutions with the boundary condition in which the value in $n\to\infty$ and that in $n\to-\infty$ differs from each other.  

\section{Discrete KdV equation and BBS}
\subsection{Ultradiscretization}

The discrete KdV equation is defined as
\begin{equation}
\frac{1}{w_{n+1}^{t+1}}-\frac{1}{w_n^t}+\frac{\delta}{1+\delta}(w_{n}^{t+1}-w_{n+1}^t)=0, \label{dkdvintro}
\end{equation}
where $n$ and $t$ take only integer values.
The discrete KdV equation is transformed to the bilinear form
\begin{equation}
(1+\delta)\sigma_{n+1}^{t+1}\sigma_n^{t-1}=\delta \sigma_{n+1}^{t-1}\sigma_n^{t+1}+\sigma_n^t\sigma_{n+1}^t, \label{kdvbil}
\end{equation}
by putting
\[
w_n^t=\frac{\sigma_n^t\sigma_{n+1}^{t-1}}{\sigma_{n+1}^{t}\sigma_{n}^{t-1}}.
\]
The ultradiscretization is a limiting procedure in which the dependent variables of the discrete equations become also discretized \cite{TOKIHIRO}.
The ultradiscretization transforms discrete equations into piecewise linear equations.  
First we use the following lemma to ultradiscretize \eqref{dkdvintro}.

\begin{lemm}
Under the boundary condition $\lim_{n\to -\infty}w_n^t=1$, the discrete KdV equation \eqref{dkdvintro} is turned into

\begin{equation}
w_{n+1}^{t+1}=\left(\delta w_{n+1}^t+(1-\delta)\prod_{k=-\infty}^n\frac{w_k^{t+1}}{w_k^t}\right)^{-1}. \label{dkdvperm}
\end{equation}

\end{lemm}

Then we put $w_n^t=\exp \left(\frac{U_n^t}{\varepsilon}\right),\ \delta=\exp\left(-\frac{L}{\varepsilon}\right)\ (\varepsilon>0)$ and take the limit $\varepsilon\to+0$ to obtain

\begin{eqnarray}
U_{n+1}^{t+1}=\min\left(L-U_{n+1}^t,\sum_{k=-\infty}^n(U_k^{t}-U_k^{t+1})\right), \label{hakotama1}
\end{eqnarray}
under the boundary condition $\lim_{n\to -\infty}U_n^t=0$.

We define the ultradiscrete KdV equation by the equation \eqref{hakotama1} \cite{HIROTA2}. It is also equivalent to the time evolution equation of BBS with box capacity $L$ \cite{SATSUMA, TAKAHASHI2}.
Next we ultradiscretize \eqref{kdvbil}. Putting $\sigma_n^t=e^{\tau_n^t/\varepsilon}$, $\delta=e^{-L/\varepsilon}$ and taking the limit $\varepsilon\to+0$, we obtain the bilinear form of the ultradiscrete KdV equation:
\begin{equation}
\tau_{n+1}^{t+1}+\tau_n^{t-1}=\max[\tau_{n+1}^{t-1}+\tau_n^{t+1}-L,\tau_n^t+\tau_{n+1}^t]. \label{udkdvbil}
\end{equation}
The ultradiscretization preserves the solitonic nature of the continuous KdV equation.
\begin{exam}

Let $L=1$, then equation \eqref{hakotama1} is closed under $U_n^t\in\{0,1\}$.
We give an example of the time evolution below (where a dot indicates a zero).
\begin{verbatim}
t=1:111...11...1..............
t=2:...111..11..1.............
t=3:......11..11.11...........
t=4:........11..1..111........
t=5:..........11.1....111.....
\end{verbatim}
\end{exam}

\subsection{BBS with a carrier}

The time evolution of the BBS with a Carrier (BBSC) is expressed as follows \cite{TAKAHASHI}.
We prepare the ``carrier'' which can carry at most $l$ balls.
At each time step in the evolution, the carrier moves from the left  to the right.
While the carrier passes the \textit{j}-th box, the following action occurs.
Assume that the carrier carries $c\ (0\le c\le l)$ balls before it passes the \textit{j}-th box, and also assume that there are $U\ (0\le U\le L)$ balls in the \textit{j}-th box.
Then, when the carrier passes the box, the carrier puts $\min(c,\ L-U)$ balls into the box and receives $\min(U,\ l-c)$ balls from the box.
That is to say, the carrier puts as many balls into the box as possible and simultaneously obtain as many balls from the box as possible.
This rule can be expressed in the following formula
\begin{small}
\begin{equation}
U_n^{t+1}=\min\left(L-U_n^t,\sum_{k=-\infty}^{n-1}(U_k^t-U_k^{t+1})\right)+\max\left(0,\sum_{k=-\infty}^nU_k^t-\sum_{k=-\infty}^{n-1}U_k^{t+1}-l\right), \label{bbscnormal}
\end{equation}
\end{small}
under the boundary condition
\[
\lim_{n\to-\infty}U_n^t=0.
\]

\bigskip
\section{Periodic BBSC that allows negative solitons}

\subsection{Negative solitons}

Negative solitons are non-solitonic trains of negative values at a speed  of $1$. These arise for example when we put negative initial values to the BBS(C) equation.

\begin{exam}
An example of a negative soliton in a BBS with box capacity $1$ is as follows (where  a dot indicates a zero).

\begin{verbatim}
t=1: . .-1-1-1 . .-1 . . . . 1 1 . . . . .
t=2: . . .-1-1-1 . .-1 . . . . . 1 1 . . .
t=3: . . . .-1-1-1 . .-1 . . . . . . 1 1 .
\end{verbatim}
While the sequence of 1's has a speed equal to its length, the sequences of -1 have speed 1 regardless of their lengths. $\Box$

\end{exam}

The pioneering works on negative solitons and background solutions are due to Hirota \cite{HIROTA1}, and Willox et.al. \cite{WILLOX}.
In the previous work with Mada and Tokihiro \cite{kanki}, the author presented a way to construct the conserved quantities of equation \eqref{hakotama1} which can also be applied to the negative solitons, by using certain gauge transformation 
to the BBSC.  We described the structure of the conserved quantities for the BBSC, in terms of arclines connecting balls and vacant boxes.

\subsection{Coupled form of BBS(C)}

Hereafter we consider the BBSC with box capacity $L$ and carrier capacity $l$.
\begin{lemm}

Equation \eqref{bbscnormal} is equivalent to the following coupled equations 
\begin{equation}
\begin{cases}
U_i^{t+1}&=\min (c_i^t,L-U_i^t)+\max (0,U_i^t+c_i^t-l), \\
c_{i+1}^t&=U_i^t+c_i^t-U_i^{t+1}, \label{bbsc}
\end{cases}
\end{equation}
with the boundary condition

\[
\lim_{i\to-\infty}c_i^t=\lim_{i\to-\infty}U_i^t=0.
\]
\end{lemm}
\begin{flushleft}\textbf{Proof}\end{flushleft}
If we determine $c_i^t$ by $c_i^t=\sum_{k=-\infty}^{i-1}(U_k^t-U_k^{t+1})$ in \eqref{bbscnormal}, we have \eqref{bbsc}. $\Box$

On the contrary, we have $c_i^t=\sum_{k=-\infty}^{i-1}(U_k^t-U_k^{t+1})$ from the last equation in \eqref{bbsc} with the boundary condition.
Then the first equation gives \eqref{bbscnormal}.
\begin{rema}
The equation \eqref{bbsc} is also called an ``ultradiscrete Yang-Baxter map''.
\end{rema}
\begin{coro}
The coupled equations of the BBS are expressed as

\begin{equation}
\begin{cases}
U_i^{t+1}&=\min (c_i^t,L-U_i^t), \\
c_{i+1}^t&=U_i^t+c_i^t-U_i^{t+1}. \label{bbslat}
\end{cases}
\end{equation}

\end{coro}
\begin{flushleft}\textbf{Proof}\end{flushleft}
We have only to set $l=+\infty$ in \eqref{bbsc}. $\Box$

\begin{figure}
\centering
\includegraphics[width=10cm, bb=100 600 400 760]{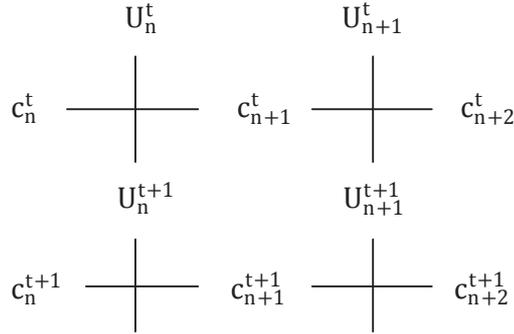}
\caption{The time evolution of the coupled BBSC equation}
\end{figure}

\subsection{Periodic BBS(C)}

We consider the time evolution of the BBS(C) with periodic boundary condition.
For a detailed discussion of the periodic BBS with $U_i,c_i$ being non-negative, see \cite{YURA} and \cite{MADA3}. We extend the results in \cite{YURA,MADA3} to the periodic BBSC that allows negative solitons.

\begin{defi}

We say that the set of initial values $U_1^t,U_2^t,\cdots U_N^t$ 
evolves as a periodic BBSC of size N,
if and only if there exists at least one $c_1^t$ such that $\ c_{N+1}^t=c_1^t$ and such that $U_1^{t+1},U_2^{t+1},\cdots U_N^{t+1}$ 
are defined uniquely, independent of the choice of possible $c_1^t$'s. $\Box$

\end{defi}
\begin{rema}
For some initial values $\{U_i^t\}_{i=1}^N$, there may exist more than one $c_1^t$'s such that $c_{N+1}^t=c_1^t$ and such that $\{U_i^{t+1}\}$ depend on $c_1^t$ (Example \ref{bbscexample}). The time evolution of the BBSC is not unique in these cases, and we exclude such cases for simplicity in this paper.
\end{rema}
\subsubsection{Periodic BBS}

First, we deal with the periodic BBS \eqref{bbslat}. 

\begin{figure}
\centering
\includegraphics[width=8cm, bb=100 620 330 800]{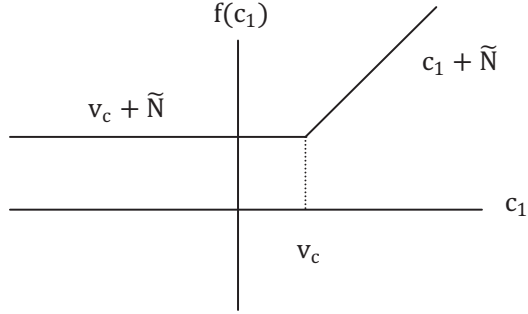}
\caption{The graph of $c_{N+1}=f(c_1)$}
\label{fc}
\end{figure}

\begin{lemm}\label{bbslem1}

\begin{equation*}
c_{N+1}^t=\max[c_1^t,v_c^t]+\tilde{N}
\end{equation*}
Here, 
\begin{eqnarray*}
v_c^t&=&\max_{i\in\{1,2,\cdots ,N\}}[iL-2(U_1^t+\cdots+U_{i-1}^t)-U_i^t],\\
\tilde{N}&=&2(U_1^t+U_2^t+\cdots U_N^t)-NL.
\end{eqnarray*}
(See figure \ref{fc}.) 
\end{lemm}

\begin{flushleft}\textbf{Proof}\end{flushleft}
First note that $\tilde{N}$ does not change under the time evolution of the periodic BBSC.
We omit the superscript $\ ({\cdot})^t$ of $c_i^t$'s and $v_c^t$ for convenience and
 consider $c_{N+1}=:f(c_1)$ as a function of $c_1$.

Fix one $c_1$ such that $c_1\ge v_c$.

Since $c_1\ge v_c\ge L-U_1$,

\begin{eqnarray*}
U_1^{t+1}&=&\min(c_1,L-U_1)=L-U_1,\\
c_2&=&c_1+2U_1-L.
\end{eqnarray*}

The inequality $c_1\ge v_c\ge 2L-2U_1-U_2$ gives

\begin{eqnarray*}
U_2^{t+1}&=&L-U_2,\\
c_3&=&c_2+2(U_1+U_2)-2L.
\end{eqnarray*}

Using $v_c\ge iL-2(U_1+\cdots+U_{i-1})-U_i$, repeated calculations lead to 

\begin{eqnarray*}
U_i^{t+1}&=&L-U_i,\\
c_{i+1}&=&c_i+2(U_1+U_2+\cdots +U_i)-iL.
\end{eqnarray*}

for $i=1,2,\cdots ,N$. (This is easily seen by induction.)

Thus we obtain $c_{N+1}=f(c_1)=c_1+\tilde{N}$ if $c_1\ge v_c$.

If $c_1< v_c$, on the other hand,
there exists $1\le J\le N$ such that

\begin{eqnarray*}
U_J^{t+1}&=&c_J,\\
c_{J+1}&=&U_J+c_J-U_J^{t+1}=U_J.
\end{eqnarray*}

holds.
Hence $(U_i,c_i)$ does not depend on $c_1$ if $i\ge J+1$. In fact $(U_i,c_i)$  will be a constant that only depends on the initial values $(U_1,\cdots ,U_N)$. 
Therefore, $c_{N+1}=f(c_1)$ is also a constant if $c_1<v_c$.
It is easily seen from \eqref{bbslat} that $f(c_1+1)-f(c_1)=0\ \text{or}\ 1$, and 
we already know that $f(c_1+1)-f(c_1)=1$ is equivalent to $c_1\ge v_c$.
Thus we obtain $f(c_1)=v_c+\tilde{N}$ if $c_1< v_c$. $\Box$

\begin{theo}\label{bbsbunrui}
The BBS \eqref{bbslat} evolves as a periodic system if and only if

\[
\tilde{N}\le 0
\]
holds for the initial values $\{U_i^0\}_{i=1}^N$.

\end{theo}

\begin{flushleft}\textbf{Proof}\end{flushleft}

From the lemma \ref{bbslem1}

\begin{itemize}
\item If $\tilde{N}>0$, no $c_1^0\in\mathbb{Z}$ satisfies $f(c_1^0)=c_1^0$.
\item If $\tilde{N}=0$, we have $f(c_1^0)=c_1^0$ for all $c_1^0\ge v_c$.
\item If $\tilde{N}<0$, the only $c_1^0$ that satisfies $f(c_1^0)=c_1^0$ is $c_1^0=v_c+\tilde{N}$. 
\end{itemize}

In the second case, $U_i^1$ is independent of the choice of $c_1^0$ because we know from the previous lemma that $U_i^1=1-U_i^0\ (1\le i\le N)$ for any  $c_1^0\ge v_c$. The evolutions are the same in $t=2,3,\cdots$. \ $\Box$

\begin{rema}

Although normally we treat the case where $U_i^t,\ c_i^t\in\mathbb{Z}$, this proposition 
is also valid for arbitrary real values of $U_i^t,\ c_i^t$. $\Box$

\end{rema}

\subsubsection{Periodic BBSC}

Next we deal with the periodic BBSC \eqref{bbsc}.
As we are investigating whether the system evolves from the time $t$ to $t+1$, we can omit the superscript $\ ({\cdot})^t$ of the variables below.

\begin{theo}\label{pbbsccomplete}
Let,
\begin{eqnarray*}
M&=&2(U_1+\cdots+U_N)-NL,\\
c_L&=&\max_{i\in\{0,1,\cdots N-1\}}[L-2(U_1+\cdots +U_{i})-U_{i+1}+iL],\\
c_R&=&\min_{i\in\{0,1,\cdots N-1\}}[l-2(U_1+\cdots +U_{i})-U_{i+1}+iL],\\
N_{\text{odd}}&=&\{i|\ 1\le i\le N,\ i:\text{odd}\},\notag\\
N_{\text{even}}&=&\{i|\ 1\le i\le N,\ i:\text{even}\},\notag\\
d_L&=&\max\left[\max_{i\in N_{\text{odd}}}(l-U_i),\ \max_{i\in N_{\text{even}}}(U_i-L+l+1)\right],\\
d_R&=&\min\left[\min_{i\in N_{\text{odd}}}(L-U_i-1),\ \min_{i\in N_{\text{even}}}(U_i)\right]+1.
\end{eqnarray*}
When we put $c_{N+1}=f(c_1)$, the following cases occur.

\begin{itemize}

\item If $\infty>l>L$:
\begin{itemize}
\item If $M\neq 0$, there exists a unique $c_1\in\mathbb{Z}$ such that $\ f(c_1)=c_1$ holds.

\item If $M=0$ and $c_L< c_R$, we have $f(c_1)=c_1$ for all $c_1$ with $c_L\le c_1\le c_R$.

\item If $M=0$ and $c_L\ge c_R$, there exists a unique $c_1$ such that $\ f(c_1)=c_1$ holds.
\end{itemize}

\item If $l<L$:
\begin{itemize}
\item If $N$ is an odd number:

There exists a unique $c_1\in\frac{1}{2}\mathbb{Z}$ such that $f(c_1)=c_1$ holds.

(For $c_1\in \frac{1}{2}\mathbb{Z}\setminus\mathbb{Z}$, we do not have a normal box ball interpretation. However, $\forall i\  U_i^t\in \mathbb{Z} \to \forall i\  U_i^{t+1}\in \mathbb{Z}$ holds, which means the number of balls in the box is an integer all the same.) $\cdots (**)$

\item If $N$ is an even number,
\begin{itemize}
\item For $d_L\ge d_R$, there is a unique $c_1\in\mathbb{Z}$ that satisfy $f(c_1)=c_1$.

\item For $d_L<d_R$, we have $f(c_1)=c_1$ for all $c_1$ such that $d_L\le c_1\le d_R$, but $\{U_i^{t+1}\}$ depends on $c_1$.

(This is the only case where we cannot determine the time evolution uniquely.) $\cdots (***)$
\end{itemize}
\end{itemize}
\item If $l=L$, there is a unique $c_1$ that satisfies $f(c_1)=c_1$.
\end{itemize}

$\Box$
\end{theo}

\begin{flushleft}\textbf{Proof}\end{flushleft}

\begin{flushleft}\textbf{\ If $l>L$ :}\end{flushleft}
We regard $c_{i+1}$ as a function of $c_i$ to find

\begin{equation}
c_{i+1}(c_i+1)-c_{i+1}(c_i)=
\begin{cases}
1\ &(L-U_i\le c_i<l-U_i),\\
0\ &(\text{otherwise}),
\end{cases}\notag
\end{equation}
which leads to

\begin{equation}
f(c_1+1)-f(c_1)=
\begin{cases}
1\ &(\forall i,\ L-U_i\le c_i<l-U_i),\\
0\ &(\text{otherwise}).
\end{cases}\notag
\end{equation}
If $L-U_i\le c_i<l-U_i$ for all \textit{i}'s then, by induction we have

\[
c_{i+1}=c_1+2(U_1+\cdots U_i)-iL\ \ (i=1,\cdots ,N).
\]
Thus the condition $(\forall i,\ L-U_i\le c_i<l-U_i)$ is equivalent to the following:

\begin{equation}
(\forall i)\ L-2(U_1+\cdots +U_i)-U_{i+1}+iL\le c_1<l-2(U_1+\cdots+ U_i)-U_{i+1}+iL. \label{clcreval}
\end{equation}
The inequality \eqref{clcreval} can be expressed as

\[
c_L\le c_1 \ \text{and}\ c_1<c_R,
\]
and we have

\begin{equation}
c_{N+1}=f(c_1)=
\begin{cases}
f(c_L)\ &(c_1\le c_L),\\
c_1+M\ &(c_L\le c_1\  \text{and}\ c_1\le c_R),\\
f(c_R)\ &(c_R\le c_1).
\end{cases}\notag
\end{equation}
(Note that $c_L$ may be larger than $c_R$.)

By examining the intersection of $y=f(c_1)$ and $y=c_1$ , we obtain the desired result.
(When $M=0$ and $c_L<c_R$, possible $c_1$'s are not unique, but we have $U_i^{t+1}=L-U_i^t$ regardless of the choice of $c_1$ which makes the time evolution of $U_i^t$ unique.)

We conclude from the above argument that the system uniquely evolves in time if $l>L$.

\begin{figure}
\centering
\includegraphics[width=10cm]{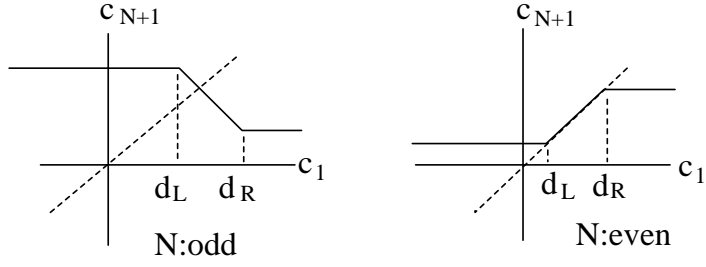}
\caption{Graph of $c_{N+1}=f(c_1)$ of BBSC under the condition $l<L$}
\label{fc2}
\end{figure}

\begin{flushleft}\textbf{\ If $l<L$ :}\end{flushleft}
We have
\begin{equation*}
c_{i+1}(c_i+1)-c_{i+1}(c_i)=
\begin{cases}
-1\ &(l-U_i\le c_i<L-U_i),\\
0\ &(\text{otherwise}).
\end{cases}
\end{equation*}
Hence,

\begin{equation*}
f(c_1+1)-f(c_1)=
\begin{cases}
1\ &(\forall i,\ l-U_i\le c_i<L-U_i)\ \text{and}\ (N:\text{even}),\\
-1\ &(\forall i,\ l-U_i\le c_i<L-U_i)\ \text{and}\ (N:\text{odd}),\\
0\ &(\text{otherwise}).
\end{cases}
\end{equation*}
If $l-U_i\le c_i<L-U_i$ for all $i$, we have

\begin{eqnarray*}
U_i^{t+1}&=&2c_i+U_i^t-l,\\
c_{i+1}&=&l-c_i,
\end{eqnarray*}
for each $i$. Thus we have
\begin{equation*}
c_i=
\begin{cases}
c_1\ &(i\in N_{\text{odd}}),\\
l-c_1\ &(i\in N_{\text{even}}).
\end{cases}
\end{equation*}
Therefore, the condition ($\forall i,\ l-U_i\le c_i<L-U_i$) is equivalent to the following set of inequalities:

\begin{equation*}
(\forall i)\ 
\begin{cases}
l-U_i\le c_1<L-U_i\ &(i\in N_{\text{odd}}),\\
U_i-L+l<c_1\le U_i\ &(i\in N_{\text{even}}).
\end{cases}
\end{equation*}
Thus if $N$ is an odd number we obtain

\begin{equation*}
f(c_1)=
\begin{cases}
f(d_L)\ &(c_1< d_L),\\
l-c_1\ &(d_L\le c_1\ \text{and}\ c_1\le d_R),\\
f(d_R)\ &(d_R< c_1).
\end{cases}
\end{equation*}
(See the left graph of figure \ref{fc2}.)

Therefore, there is a unique $c_1$ such that $f(c_1)=c_1$ ($c_1\in\frac{1}{2}\mathbb{Z}$).
Even though $c_1$ may not be an integer, as long as $\{U_i^t\}$ are all integers, $\{U_i^{t+1}\}$ are closed in $\mathbb{Z}$.
In particular if $d_L\ge d_R$, $f(c_1)$  is  constant in $c_1$.

If $N$ is an even number, we obtain:
\begin{equation*}
f(c_1)=
\begin{cases}
f(d_L)\ &(d_1< d_L),\\
c_1\ &(d_L\le c_1\ \text{and}\ c_1\le d_R),\\
f(d_R)\ &(d_R< c_1).
\end{cases}
\end{equation*}
(See the right graph of figure \ref{fc2}.)

Therefore if $d_L\ge d_R$,  $f(c_1)$ is constant in $c_1$, which renders the time evolution unique.

If $d_L<d_R$, by fixing $c_1$ such that $d_L\le c_1\le d_R$ we obtain

\[
U_i^{t+1}=U_i^t+(-1)^i(l-2c_1).
\]
This indicates that the time evolution is dependent on $c_1$. We do not have a unique periodic BBSC in this case.$\ \Box$

\begin{coro}
The time evolution of the BBSC \eqref{bbsc} with the periodic boundary condition is \underbar{not} well-defined if and only if the three conditions
\begin{equation*}
\begin{cases}
l<L,\\
\text{the system size}\ N \ \text{is an even number},\\
d_L<d_R, 
\end{cases}
\end{equation*}
hold simultaneously. 

\end{coro}

\begin{exam}
We show an example of the periodic BBSC. We express the time evolution 
\begin{eqnarray*}
\begin{cases}
V=\min(c,L-U)+\max(0,U+c-l),\\
d=U+c-V,
\end{cases}
\end{eqnarray*}
as follows:\begin{verbatim}
  U
c-|-d.
  V
\end{verbatim}
Let $N=8$, $L=1$ and $l\gg 1$.
\begin{verbatim}
  0   3   0   0   0   0   0   0
0-|-0-|-5-|-4-|-3-|-2-|-1-|-0-|-0
  0   a   1   1   1   1   1   0
4-|-3-|-a-|-1-|-2-|-3-|-4-|-5-|-4
  1   3   a   0   0   0   0   1
1-|-2-|-7-|-2-|-1-|-0-|-0-|-0-|-1
  0   a   3   1   1   0   0   0
2-|-1-|-a-|-3-|-4-|-5-|-4-|-3-|-2
  1   1   a   0   0   1   1   1
3-|-4-|-5-|-0-|-0-|-0-|-1-|-2-|-3
  0   0   3   0   0   0   0   0
\end{verbatim}
Here $a=-2$.
\end{exam}
\begin{exam}\label{bbscexample}
We present two irregular cases in defining periodic BBSC.
Here is an example of the case $(**)$ in proposition \ref{pbbsccomplete}. Let $N=7$, $L=5$, $l=1$ and $b=\frac{1}{2}$.
\begin{verbatim}
  1   4   1   4   1   4   1
b-|-b-|-b-|-b-|-b-|-b-|-b-|-b
  1   4   1   4   1   4   1
\end{verbatim}
Next we show an example when the periodic BBSC is not well-defined.
Let $N=4$, $L=5$ and $l=2$.
\begin{verbatim}
  2   3   2   3          2   3   2   3
1-|-1-|-1-|-1-|-1 ,    2-|-0-|-2-|-0-|-2
  2   3   2   3          4   1   4   1
\end{verbatim}
We have more than one types of time evolutions depending on the choice of $c_1$.
\end{exam}

\begin{rema}
The BBSC is an invertible system, which is true even when some of the variables $\{U_i^t,c_i^t\}$ take negative values. 
\end{rema}
\subsection{Periodic BBS with $K$ kinds of balls}

\begin{figure}
\centering
\includegraphics[width=10cm, bb=150 650 500 750]{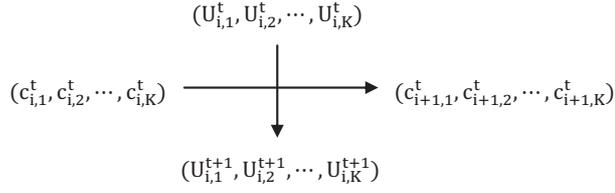}
\caption{Time evolution of BBS$_K$}
\label{bbsk}
\end{figure}

We consider the extended BBS where we have $K$ kinds of balls distinguished by integers $1\le k\le K$. We call this system BBS$_K$. Here we consider the case where the capacity of the box is $1$ in each box.
Let $U_{i,k}^t$ be the number of balls $k$ in the $i$-th box and $c_{i,k}^t$ the number of balls $k$ in the carrier when the carrier is located between the $(i-1)$-th box and the $i$-th one.
The time evolution rule of the BBS$_K$ is
\begin{eqnarray*}
U_{i,k}^t&=&\min\left(1-\sum_{j=1}^{k-1}U_{i,j}^{t+1}-\sum_{j=k}^K U_{i,j}^t,\ c_{i,k}^t\right),\\
c_{i+1,k}^t&=&U_{i,k}^t+c_{i,k}^t-U_{i,k}^{t+1},
\end{eqnarray*}
where $k=1,2,\cdots ,K$ (See figure \ref{bbsk}).

\begin{defi}
The BBS$_K$ can be defined as a periodic system if there exists at least one set of \[(c_{1,1}^t,c_{1,2}^t, \cdots ,c_{1,K}^t)\in\mathbb{R}^K\] 
such that $c_{N+1,j}^t=c_{1,j}^t$ hold for all $j\in\{1,\cdots ,K\}$ and the $U_{i,k}^{t+1}$'s are determined uniquely independent of the choice of possible $c_{1,k}^t$'s.
\end{defi}

\begin{theo}
The periodic BBS$_K$ is well-defined if and only if 
\[
\sum_{j=1}^N U_{j,l}^0\le \frac{1}{2}\left(N-\sum_{j=1}^N\sum_{\genfrac{}{}{0pt}{}{k=1}{k\neq l}}^K U_{j,k}^0\right)
\]
holds for all $l\in\{1,2,\cdots ,N\}$.
\end{theo}

\textbf{Proof}

First note that we only have to consider the case where $t=0$ because $\sum_{j=1}^N U_{j,l}^t$ is independent of $t$.
We know from the time evolution rule that $c_{N+1,k}^0$ depends only on $c_{1,1}^0,c_{1,2}^0,\cdots ,c_{1,k}^0$ and does not depend on 
$c_{1,k+1}^0,c_{1,k+2}^0,\cdots , c_{1,K}^0$. We omit the superscript ${(\cdot)}^0$ for convenience hereafter.

For $l=1,2,\cdots , N$ let
\[
N_l=\sum_{j=1}^N\sum_{k=1}^K U_{j,k}+\sum_{j=1}^N U_{j,l}-N.
\]
The condition in the proposition is rewritten as $N_l\le 0$ for all $l$.

We first determine the cases that allow for a $c_{1,1}$ such that satisfies $c_{N+1,1}=c_{1,1}$.
We calculate in the same way as in lemma \ref{bbslem1} to obtain
\begin{eqnarray*}
c_{N+1,1}&=&\max[c_{1,1},\gamma_1]+N_1,\\
\gamma_1&=&\max_{i\in\{1,\cdots ,N\}}\left[i-\sum_{j=1}^{i-1}U_{j,1}-\sum_{j=1}^i\sum_{k=1}^K U_{j,k}\right].
\end{eqnarray*}
Thus there exists $c_{1,1}$ such that $c_{N+1,1}=c_{1,1}$ if and only if $N_1\le 0$.

Next we determine $c_{N+1,2}$ from $c_{1,1}$ and $c_{1,2}$.
We obtain
\begin{eqnarray*}
c_{N+1,2}&=&\max[c_{1,2},\gamma_2]+N'_2,\\
N'_2&=&\sum_{j=1}^N\left(\tilde{U}_{j,1}+\sum_{k=2}^K U_{j,k}\right)+\sum_{j=1}^N U_{j,2}-N,
\end{eqnarray*}
and $\gamma_2$ is a constant determined by the $U_{i,k}$'s only. 
Therefore we have $c_{1,2}$ such that $c_{N+1,2}=c_{1,2}$ if and only if $N'_2\le 0$.
We know from $\sum_{i=1}^N \tilde{U}_{i,1}=\sum_{i=1}^N U_{i,1}$ that $N'_2=N_2$.
By repeating the same calculations for $c_{N+1,k}$ $(k\ge 3)$ we obtain the proposition.
If $N_j=0$ for some $j$ there exist more than one $c_{1,j}$ such that $c_{N+1,j}=c_{1,j}$.
The time evolution of $U_{i,k}$ is still unique in this case.
$\Box$

\subsection{Some elaborations on conserved quantities}
\begin{prop}\label{sokoage}

The invertible transformation

\begin{equation*}
\begin{cases}
\tilde{U}_i^t&=U_i^t+m, \\
\tilde{c}_i^t&=c_i^t+m,
\end{cases}
\end{equation*}
deforms BBSC \eqref{bbsc} to

\begin{equation}
\begin{cases}
\tilde{U}_i^{t+1}&=\min (\tilde{c}_i^t,(L+2m)-\tilde{U}_i^t)+\max (0,\tilde{U}_i^t+\tilde{c}_i^t-(l+2m)), \\
\tilde{c}_{i+1}^t&=\tilde{U}_i^t+\tilde{c}_i^t-\tilde{U}_i^{t+1}. 
\end{cases}\label{gaugeBBSC}
\end{equation}
Here $m$ is an arbitrary real number.
\end{prop}

\begin{flushleft}\textbf{Proof}\end{flushleft}
We immediately obtain \eqref{gaugeBBSC} by direct calculation \cite{nagai}. $\Box$

When $m>0$, in particular, we call this transformation an ``upward-shift translation''.
If some of the variables take negative values in the initial conditions, by putting $m>0$ large enough, all $U_i^t$'s and $c_i^t$'s are transformed to be  
positive.
Note that the capacity of the boxes and that of the carrier increase by $2m$, and that the boundary conditions will change: $U_i^t$ is not $0$ at $i\to\pm \infty$ anymore, but it becomes some nonzero constant at $i\to\pm \infty$.
Also note that the value $[2(U_1^0+U_2^0+\cdots U_N^0)-NL]$ does not change through this shift.
Hence, treating the negative solitons is equivalent to treating the non-negative BBS's with the boundary conditions that the solutions $U_n^t$ have some positive constant value at $n\to \pm \infty$.
\begin{rema}
In this transformation, the values $U_i^t$ at $i\to+\infty$ and those at $i\to-\infty$ have to be the same. In section $5$, we present a way to deal with the boundary condition such that $\lim_{i\to-\infty}U_i^t\neq \lim_{i\to+\infty}U_i^t$.
 $\Box$
\end{rema}
The construction of conserved quantities found in \cite{kanki} can also be performed for a periodic BBSC with general $L$ and $l$.
We depict by arclines the movement of the balls from the box to the other box according to the time evolution of BBSC. The operation of drawing arclines when the capacity of the carrier is $l$ is denoted by (OP)$_l$.

\underline{\textbf{$($OP$)_l$}}
\begin{itemize}
\item 
We take out balls from the boxes from the left  to the right according to the evolution of BBSC. We distinguish each ball and when two or more balls are taken out from the same box at each step, the one at a lower position is taken out first. The carrier can hold at most $l$ balls.
\item 
When  two or more balls are passed from the carrier to a box at the same step, the one taken by the carrier last will be the first to go back to a box. 
\item 
We depict by arclines the movement of the balls from the box to the carrier to the other box.
\end{itemize}
We can see some structures in the set of arclines connecting the balls and the empty boxes. 


\begin{theo}[Kanki-Mada-Tokihiro \cite{kanki}]\label{theokmt}
On the periodic BBSC upward-shifted by $m$, let
\[
\tilde{C}_l:=\#\{\text{arclines drawn at (OP)}_l\}-\#\{\text{arclines drawn at (OP)}_{l-1}\},
\]
for the given initial condition.
Then $(\tilde{C}_1,\tilde{C}_2,\cdots)$ is a set of constants independent of the time evolution of the system.
We also denote by $(C^0_1,C^0_2,\cdots)$ the set of these constants for the ``vacuum state'' where every box has exactly $m$ balls.
The difference of these two $(\tilde{C}_1-C^0_1,\tilde{C}_2-C^0_2,\cdots)$ is also a set conserved quantities of the BBSC.
We rewrite it $(C_1,C_2,\cdots)$ and call it ``the set of conserved quantities of the BBSC''. $\Box$
\end{theo}

We extend this theorem to the general periodic BBSC. 
\begin{prop}
For the periodic BBSC with parameters $L$ and $l$ and initial conditions $\{U_i^0\}$ that satisfy the following

\begin{eqnarray}\label{prescondi}
&&\text{For both the initial conditions}\  \{U_i^0\}\ \text{and the vacuum solution}\ \{m\} \notag\\
&&\text{we can define the periodic system for}\  l\in[L,+\infty], \notag\\
&&\text{and also for all}\ l, \text{there exists} \ c_1\in\mathbb{Z} \ \text{such that} \ f(c_1)=c_1,
\end{eqnarray}
we can apply theorem \ref{theokmt} to construct the conserved quantities. $\Box$
\end{prop}

In other words, we consider the cases other than $(**)$ in proposition \ref{pbbsccomplete}. 
Note that for $l$ with $l<L$ we allow the case $(***)$ in proposition \ref{pbbsccomplete}.
\begin{exam}
See figure \ref{bbscpair} for an example. We consider the periodic BBSC with a box capacity $1$ and system size $N=12$ and set the initial value to be $001\underline{1}01101000$.
By an upward-shift with $m=1$, initial values are transformed into $112012212111$. We have $(\tilde{C}_1, \tilde{C}_2, \tilde{C}_3, \tilde{C}_4)=(6,5,3,1)$, $(C_1^0, C_2^0, C_3^0, C_4^0)=(6,6,0,0)$ and $(C_1, C_2, C_3, C_4)=(0,-1,3,1)$.
\begin{figure}
\centering
\includegraphics[width=9cm,bb=70 425 350 740]{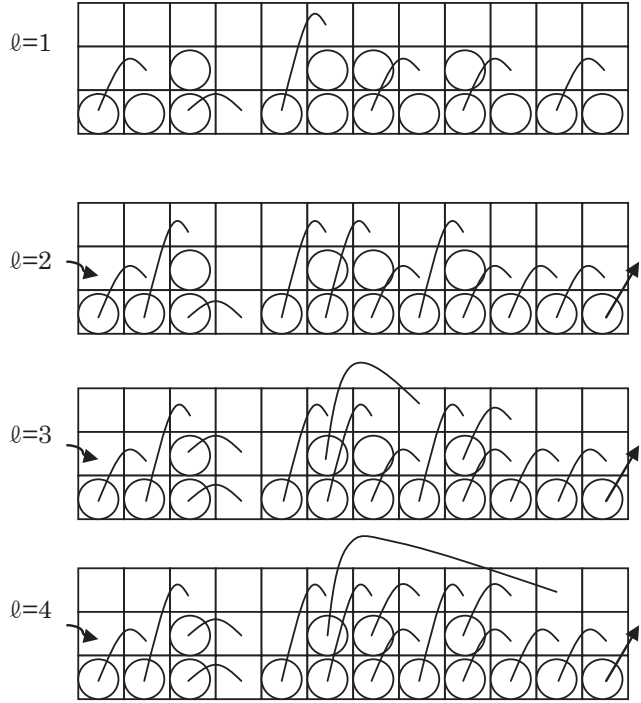}
\caption{Drawing arclines to the periodic system of size $N=12$.}
\label{bbscpair}
\end{figure}
\end{exam}
\subsection{Defining Background Solutions}

We propose the following way to distinguish the background solutions from the positive solitons.
Note that the balls connected in (OP)$_l$ are included in those connected in (OP)$_{l'}$ where $l'>l$.
\begin{defi}
For the BBS upward-shifted by $m$, the balls connected in the process $($OP)$_{2m}$ constitute negative soliton solutions. Remaining balls constitute positive solitons.
\end{defi}
\begin{defi}
The solution $\{U_i^t\}$ is a ``background solution'' if
$C_k=0$ for $\forall k\ge 2m+1$.
\end{defi}
\begin{exam}\label{bbscex22}
For the BBS with the box capacity one,
$\cdots 0\underline{1}1\underline{1}0\cdots$ and $\cdots 0\underline{1}2\underline{1}0\cdots$ are both stationary solitary waves moving at speed one.
We can distinguish these two waves by constructing conserved quantities. The former is a background solution \underline{without} positive solitons,  and the latter can be interpreted as a background solution with a soliton of length one superimposed. 
We shift both systems upward by $m=1$, and find that the former has the conserved quantity $(C_1,C_2)=(0,-1)$, $C_k=0 (k\ge 3)$. On the other hand the latter has $(C_1,C_2,C_3)=(0,-1,1)$. (See figure \ref{bbscex1})

\begin{figure}
\centering
\includegraphics[width=12cm,bb=80 620 450 770]{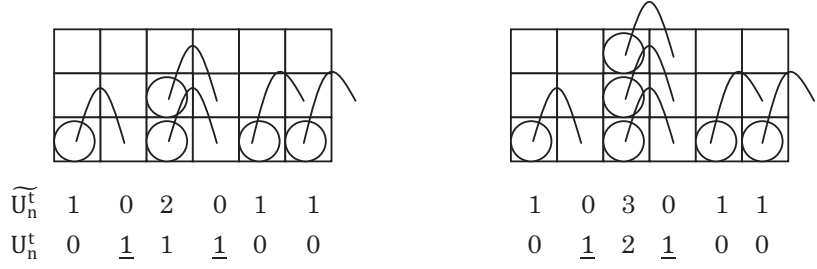}
\caption{(Example \ref{bbscex22}):Distinguishing between -11-1 and -12-1.}
\label{bbscex1}
\end{figure}
\end{exam}

\bigskip

\section{Relation to the discrete system}

We consider the following transformation
\begin{equation}
\tilde{\sigma}_n^t=\delta^{-(t-n)^2/2L}\sigma_n^t \label{dkdvtrans43}
\end{equation}
for the solution $\sigma_n^t$ of the bilinear discrete KdV equation.
\begin{prop}
The transformed function $\tilde{\sigma}_n^t$ satisfies the following gauge transformed discrete KdV equation

\begin{equation}
(1+\delta)\tilde{\sigma}_{n+1}^{t+1}\tilde{\sigma}_{n}^{t-1}=\delta^{(1+2/L)}\tilde
{\sigma}_{n+1}^{t-1}\tilde{\sigma}_{n}^{t+1}+\tilde{\sigma}_n^t\tilde{\sigma}_{n+1}^t.\ \Box \label{gaugedkdv1873}
\end{equation}
\end{prop}
This statement is proved by a direct calculation. We consider the ultradiscrete limit of the equation \eqref{gaugedkdv1873}. We let $\delta=e^{-L/\varepsilon}$, $\sigma_n^t=\text{e}^{\tau_n^t/\varepsilon}$ and $\tilde{\sigma}_n^t=e^{\tilde{\tau}_n^t/\varepsilon}$, and the following proposition holds.
\begin{prop}
For $U_n^t$ and $\tilde{U}_n^t$ defined as $U_n^t=\tau_n^t+\tau_{n-1}^{t+1}-\tau_n^{t+1}-\tau_{n-1}^t$ and
$\tilde{U}_n^t=\tilde{\tau}_n^t+\tilde{\tau}_{n-1}^{t+1}-\tilde{\tau}_n^{t+1}-\tilde{\tau}_{n-1}^t$,
the relation \eqref{dkdvtrans43} is transformed into 
\[\tilde{U}_n^t=U_n^t+1.\]
in the limit $\varepsilon\to +0$. $\Box$
\end{prop}

\begin{flushleft}\textbf{Proof}\end{flushleft}

We take the limit $\varepsilon\to +0$ to obtain
$\tilde{\tau}_n^t=\tau_n^t+\frac{1}{2}(t-n)^2$.
Therefore we have $\tilde{U}_n^t=U_n^t+1$. $\Box$
\begin{rema}
The transformation \eqref{dkdvtrans43} of the discrete KdV equation corresponds to the `\textit{1}' upward-shift to the BBSC in the ultradiscrete limit. Thus the solution of the equation \eqref{gaugedkdv1873} goes to a `\textit{1}' upward-shifted solution of the BBSC in the ultradiscrete limit. 
\end{rema}
\bigskip

\section{Background solutions from Jacobi theta functions}

We define Jacobi theta function by

\[
\vartheta_3(v)=\vartheta_3(v,\eta)=\sum_{n=-\infty}^\infty q^{n^2}z^{2n},
\]
where $\ q=e^{i\pi\eta}$, $z=e^{i\pi v}$ and $i=\sqrt{-1}$.
We suppose Im$(\eta)>0$.
The Jacobi theta functions are quasi-doubly periodic functions, that is, we have

\begin{eqnarray*}
\{\vartheta_3(v+1)\}^2&=&\{\vartheta_3(v)\}^2,\\
\{\vartheta_3(v+\eta)\}^2&=&e^{-2\pi i(2v+\eta)}\{\vartheta_3(v)\}^2.
\end{eqnarray*}

\begin{prop}

Let $v=\eta(n_1-n_2-n_3)+\eta_0$. Then the function $\tau(n_1,n_2,n_3) :=\vartheta_3(v)$ is a solution of the following equation, which is a gauge-transformed discrete KP equation

\begin{equation}
(s\alpha)\tau_1\tau_{23}-\tau_2\tau_{13}+(1-s)\tau_3\tau_{12}=0. \label{kplike}
\end{equation}

Here, $\alpha=e^{4i\pi\eta}(=q^4)$ and $s$ is an arbitrary complex number. The lower script `$i$' of the function $\tau$ denotes a `$+1$' shift in the variable $n_i$, i.e., $\tau_1=\tau(n_1+1, n_2, n_3)$, $\tau_{23}=\tau(n_1, n_2+1, n_3+1)$, etc $\cdots$.

\end{prop}

\begin{flushleft}\textbf{Proof}\end{flushleft}

From the quasi-doubly periodicity we have
\begin{eqnarray*}
\tau_1&=&\vartheta_3(v+\eta )=e^{-i\pi(2v+\eta )}\vartheta_3(v),\\
\tau_{23}&=&\vartheta_3(v-2\eta)=e^{i\pi(4v-4\eta)}\vartheta_3(v),
\end{eqnarray*}
which lead to

\begin{equation}
\tau_1\tau_{23}=e^{i\pi(2v-5\eta)}(\vartheta_3(v))^2. \label{teta1}
\end{equation}

We also obtain
\begin{eqnarray}
\tau_2\tau_{13}=e^{i\pi(2v-\eta)}(\vartheta_3(v))^2, \label{teta2}\\
\tau_3\tau_{12}=e^{i\pi(2v-\eta)}(\vartheta_3(v))^2. \label{teta3}
\end{eqnarray}

Thus we have the result. $\Box$

\begin{rema}
If a function $\alpha(n_1,n_2,n_3)$ satisfies the relation
$\alpha\tau_1\tau_{23}=\tau_2\tau_{13}=\tau_3\tau_{12}$,
then $\tau(n_1,n_2,n_3)$ satisfies the same equation \eqref{kplike}.
The discussions in this section are therefore equally valid for such functions $\alpha$.
\end{rema}

\subsection{Reduction}

Since the parameter $s$ is arbitrary, we can put $s$ to be $s=e^{i\pi\eta\cdot \xi}\ (\xi>0)$.
The parameter $\xi$ is related to the box capacity of the BBS.
From the way $v$ is chosen, we have $\tau=\tau_{12}$, which is the 
reduction condition from the KP equation to the KdV equation. 
If we rewrite $\tau_n^t:=\tau(t,0,n)$ the following lemma holds.

\begin{lemm}
The gauge-transformed discrete KP equation \eqref{kplike} is reduced to the following 
gauge-transformed discrete KdV equation

\begin{equation}
s\alpha\tau_n^{t+1}\tau_{n+1}^{t-1}+(1-s)\tau_{n+1}^t \tau_n^t=\tau_{n}^{t-1}\tau_{n+1}^{t+1}. \label{dkdv1}
\end{equation}
\end{lemm}

\begin{rema}
We have the following relation with the upward shifted discrete KdV equation \eqref{gaugedkdv1873}:
\begin{eqnarray*}
s&=&\frac{\delta}{1+\delta},\\
\alpha&=&\delta^{2/L}. \ \ \ \ \ \Box
\end{eqnarray*}
\end{rema}
Although the coefficients in equation \eqref{dkdv1} are different from the bilinear form of the normal discrete KdV equation, we find from the following proposition that this does not lead to the loss of generality.

\begin{prop}[Tsujimoto-Hirota \cite{HIROTA2}] \label{kdvbilinear}
By change of variable
\[
w_n^t=\frac{\tau_n^t\tau_{n+1}^{t-1}}{\tau_{n+1}^t\tau_{n}^{t-1}},
\]
the equation \eqref{dkdv1} is transformed into the discrete KdV equation

\[
\frac{1}{w_{n+1}^{t+1}}-\frac{1}{w_n^t}=\delta'(w_{n+1}^t-w_n^{t+1}).
\]
where $\delta'=s\alpha$.
\end{prop}

\begin{flushleft}\textbf{Proof}\end{flushleft}

From the following identity
\begin{align*}
&(\text{LHS of \eqref{dkdv1}})_{n\to n+1}\times\tau_n^t-(\text{LHS of \eqref{dkdv1}})\times\tau_{n+2}^t\\
&=(\text{RHS of \eqref{dkdv1}})_{n\to n+1}\times\tau_n^t-(\text{RHS of \eqref{dkdv1}})\times\tau_{n+2}^t,
\end{align*}
we have
\[
\tau_{n+2}^{t+1}\tau_n^t\tau_{n+1}^{t-1}-\tau_{n+1}^{t+1}\tau_{n+2}^{t}\tau_n^{t-1}=
\delta'\tau_n^t\tau_{n+1}^{t+1}\tau_{n+2}^{t-1}-\delta'\tau_{n+2}^{t}\tau_{n+1}^{t-1}\tau_n^{t+1}.
\]
Multiplying 

\[
\frac{\tau_{n+1}^t}{\tau_{n+1}^{t+1}\tau_{n+2}^t\tau_n^t\tau_{n+1}^{t-1}}
\]
on both sides, we have the result. $\Box$

\subsection{Coexistence of Solitons and Jacobi theta backgrounds}

If we ultradiscretize the solutions of the equation \eqref{dkdv1}, we will obtain the solutions of the BBS.
We will ultradiscretize the equation and the solutions.
To do this we first assume that $\eta$ and $\eta_0$ are both purely imaginary and then replace $i\pi\eta$ with $-L/\varepsilon$. (This choice of $\varepsilon$ is empirical.) Then we take logarithms on both sides and take the limit $\varepsilon\to +0$.

Note that our paper is not the first to ultradiscretize the theta functions. For example in \cite{IWAO}, Iwao and Tokihiro ultradiscretized the theta function solution of the periodic Toda equation. Our method is unique in that we connected the ultradiscrete theta function to the background solution of the BBS.

\begin{lemm}
The ultradiscretization of the solution $\tau_n^t=\vartheta_3(v)$ is the tau function $\Theta_n^t$ of the constant solution which takes $U_n^t=2L$ for $n\in(-\infty,\infty)$.
Here we have the following relation between $U_n^t$ and $\Theta_n^t$.
\[
U_n^t=\Theta_n^t+\Theta_{n-1}^{t+1}-\Theta_n^{t+1}-\Theta_{n-1}^t.
\]
\end{lemm}

\begin{flushleft}\textbf{Proof}\end{flushleft}

\begin{eqnarray*}
\vartheta_3(v)&=&\sum_{k=-\infty}^\infty e^{i\pi\eta k^2}e^{2i\pi k\cdot v} \notag \\
&=&\sum_{k=-\infty}^\infty \exp\left[-i\pi\eta k\{2(t-n)-k+2\eta_0/\eta\}\right].
\end{eqnarray*}
From the first line to the second line we replaced $k$ with $-k$.
We then transform $i\pi\eta$ into $-\frac{L}{\varepsilon}$, and affect on both sides  $\lim_{\varepsilon\to +0}\varepsilon \log(\cdot)$. If we write the left hand side $\Theta_n^t$ after the ultradiscretization, we have

\[
\Theta_n^t=\max_{k\in \mathbb{Z}}[Lk(2(t-n)-k+\eta')].
\]

Here we put $\eta'=2\eta_0/\eta$.
The solution $U_n^t$ constructed from $\Theta_n^t$ takes constant value $2L$ on 
$n\in(-\infty,\infty)$. $\Box$

We now have the background state $\vartheta_3(v)$.
We can add \textit{N}-soliton solutions onto the background state $\vartheta_3(v)$.
If we suppose that $\tau_n^t:=\vartheta_3(v)\phi_n^t$ is also a solution of \eqref{dkdv1}, then $\phi_n^t$ satisfies the ordinary discrete KdV equation
\begin{equation*}
s\phi_n^{t+1}\phi_{n+1}^{t-1}+(1-s)\phi_{n+1}^t \phi_n^t=\phi_{n}^{t-1}\phi_{n+1}^{t+1}.
\end{equation*}

We can take $\phi_n^t$ to be the \textit{N}-soliton solution of the discrete KdV equation.
We denote the ultradiscrete limit of $\phi_n^t$ by $\Phi_n^t$ and define $V_n^t$ by $V_n^t=\Phi_n^t+\Phi_{n-1}^{t+1}-\Phi_n^{t+1}-\Phi_{n-1}^t$.
Since $s=e^{i\pi\eta\xi}=e^{-L\xi/\varepsilon}$, the solution $V_n^t$ corresponds to the solution of BBS with box capacity $L\xi$.

\begin{prop}\label{permanentkdv}
We have the BBS with box capacity $(4+\xi)L$ from the equation \eqref{dkdv1} through ultradiscretization.
\end{prop}

\begin{flushleft}\textbf{Proof}\end{flushleft}
The proof is based on Takahashi and Hirota \cite{TAKAHIRO}.
If we put $\delta'=se^{4i\pi\eta}$, then \eqref{dkdv1} is transformed into
\[
\frac{1}{w_{n+1}^{t+1}}-\frac{1}{w_n^t}=\delta'(w_{n+1}^t-w_n^{t+1})
\]
from the proposition \ref{kdvbilinear}. Thus we have

\begin{equation}
\frac{w_{n+1}^{t+1}}{w_n^t}=\frac{1-\delta'w_{n+1}^tw_{n+1}^{t+1}}{1-\delta'w_{n}^tw_{n}^{t+1}}. \label{permanent}
\end{equation}

From \eqref{permanent} we obtain for any $M<0$
\[
\prod_{k=-M}^n \frac{w_{k+1}^{t+1}}{w_k^t}=\frac{1-\delta'w_{n+1}^tw_{n+1}^{t+1}}{1-\delta'w_{-M}^tw_{-M}^{t+1}}.
\]

Here we have
\[
w_n^t=\frac{\tau_n^t\tau_{n+1}^{t+1}}{\tau_{n+1}^t\tau_{n}^{t-1}}=\frac{\vartheta_n^t\vartheta_{n+1}^{t+1}}{\vartheta_{n+1}^t\vartheta_{n}^{t-1}}
\frac{\phi_n^t\phi_{n+1}^{t+1}}{\phi_{n+1}^t\phi_{n}^{t-1}},
\]
where 
\[
\frac{\vartheta_n^t\vartheta_{n+1}^{t+1}}{\vartheta_{n+1}^t\vartheta_{n}^{t-1}}=\frac{\vartheta_3(v)\vartheta_3(v-2\eta)}{(\vartheta_3(v-\eta))^2}=e^{-2i\pi\eta}.
\]
and the positive solitons satisfy $\lim_{n\to-\infty}\phi_n^t=1$.

Thus we obtain $\lim_{n\to -\infty}w_n^t=e^{-2i\pi\eta}$.
Therefore in the limit $M\to+\infty$ we have
\[
\prod_{k=-\infty}^n \frac{w_{k+1}^{t+1}}{w_k^t}=\frac{1}{1-s}(1-\delta'w_{n+1}^tw_{n+1}^{t+1}),
\]
which is equivalent to
\[
w_{n+1}^{t+1}=\left(\delta'w_{n+1}^t+(1-s)\prod_{k=-\infty}^n \frac{w_{k}^{t+1}}{w_k^t}\right)^{-1}.
\]

After replacing $i\pi\eta$ with $-\frac{L}{\varepsilon}$, and $w_n^t$ with $e^{U_n^t/\varepsilon}$, we take the limit $\varepsilon\to 0$ to obtain

\[
U_{n+1}^{t+1}=-\max[U_{n+1}^t-(4+\xi)L,\ \sum_{k=-\infty}^n(U_k^{t+1}-U_k^t)].
\]

From $-\max(a,b)=\min(-a,-b)$, the equation is the BBS with the box capacity $(4+\xi)L$. $\Box$

These results can be summed up to the following theorem.


\begin{theo}

As solutions of the gauge-transformed discrete KdV equation \eqref{dkdv1}, we have the following type of solutions $\tau_n^t$ constructed from the 
Jacobi theta functions.

\[
\tau_n^t=\underbrace{\vartheta_3(v)}_{(\text{Background solution})} \times \underbrace{\phi_n^t}_{(\text{N-soliton solution})}.
\]

The ultradiscretization of this solution is a solution of the BBS with box capacity $(4+\xi)L$ and is $2L+V_n^t$ .
$\Box$
\end{theo}
\begin{exam}
For example when $L=1/2, \xi=2$, we have a solution of the BBS with box capacity $3$ as shown in figure \ref{tetasolex1}.

\begin{figure}
\centering
\includegraphics[width=10cm, bb=80 560 440 790]{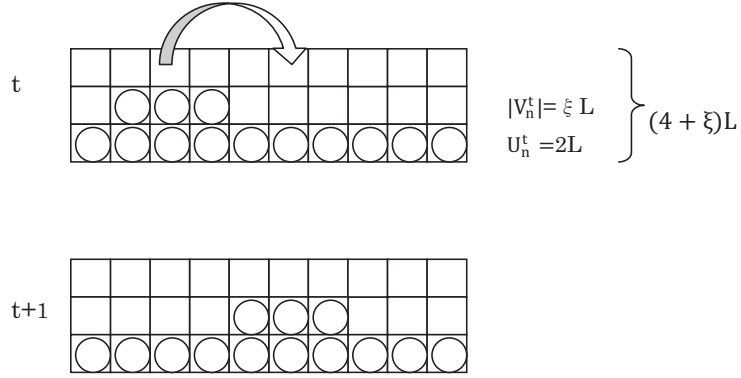}
\caption{Example of background and \textit{1}-soliton solution constructed from Jacobi theta function}
\label{tetasolex1}
\end{figure}
\end{exam}
We successfully obtained the positive solitons $V_n^t$ superimposed over the background solutions $U_n^t$. However, to obtain negative solitons we have to consider solutions with holes in the background solutions, which needs a more detailed discussion \cite{kanki}.

\bigskip

\section{BBS with irregular boundary conditions}

Finally we present the BBS with a boundary condition where 
\[\lim_{n\to+\infty} U_n^t\neq \lim_{n\to-\infty} U_n^t.\]
\begin{lemm}
The $\tau$-function of the N-soliton solution of the BBS is as follows
\begin{small}
\begin{equation}
\tau_n^t=\max_{J\subset [N]}\left[\left(\sum_{i\in J}P_i\right)t-\left(\sum_{i\in J}\min (L,P_i)\right)n+\left(\sum_{i\in J}\theta_i\right)-\sum_{i>j\ i,j\in J}2\min(P_i,P_j)\right], \label{bbssol}
\end{equation}
\end{small}
where $\forall i\ P_i>0$ and $L_i\in\mathbb{R}$.
\end{lemm}

We take $L=3$ and consider the soliton solution made up of solitons $(P_i,\theta_i)=(2,1)\ (i=1,2,\cdots ,N_0)$. It has the expression

\[
\tau(x)=\max_{0\le |J|\le N_0}\left[2|J|x+|J|-2|J|(|J|-1)\right].\ \ (x=t-n)
\]
In the limit $N_0\to \infty$, the $N_0$-soliton solution converges to the following background solution:

\begin{equation}
\tau_{bg}(x)=\max_{k\ge 0}\left[2k\left(x+\frac{1}{2}\right)-2k(k-1)\right]. \label{irregularbk}
\end{equation}

Values of $U_n^t=\tau_{bg}(x)+\tau_{bg}(x+2)-2\tau_{bg}(x+1)$ at integer points at time $t=0$ are as follows.

\begin{tabular}{|l|clll|lllc|} \hline
$n$ & $\cdots$ & 0 & 1 & 2 & 3 & 4 & 5 & $\cdots$\\ \hline
$U_n^t$ & 1 & 1 & 1 & 1 & 0 & 0 & 0 & 0\\ \hline
\end{tabular}

The background solution \eqref{irregularbk} has the new boundary condition where the solution tends to  $1$ in $n\to -\infty$ and  to $0$ in $n\to\infty$. 
We can add positive solitons to \eqref{irregularbk}.

\begin{exam}

We construct the following 2-soliton solution travelling in the background state \eqref{irregularbk}.
Let the width and the phases of the two solitons to be added be

\[
\hat{P}_1=4, \hat{\theta}_1=-8;\ \hat{P}_2=5, \hat{\theta}_2=-12,
\]
and we denote these solitons by soliton $A$ and $B$ respectively. The general form of the solution $\tau(x)$ is given as
\begin{align*}
\tau(x)&=\max[\tau_{bg}(x),\ 4t-3n-8+\tau_{bg}(x-2),\\ &5t-3n-12+\tau_{bg}(x-2),\ 9t-6n-28+\tau_{bg}(x-4)].
\end{align*}
Construction of the coexisting state of solitons and negative solitons is based on \cite{kanki}.
What happens during the time evolution of this system is as follows.
(See also the time evolution pattern of the system at the end of this section.)
\begin{itemize}
\item $0\le t\le 5$

We observe that the soliton $A$ travels at speed $2$ and the soliton $B$ travels at speed $3$ on the background $1$. The soliton $B$ takes over the soliton $A$ in the same way as in BBS with box capacity $1$.

\item $t \sim 6$

The soliton $B$ climbs down the discontinuity of the background state from $1$ to $0$. The phase of the background is shifted $2$ to the left. The soliton $B$ travels at speed $5/3$ from now on.

\item $t \sim 12$

The same incident happens to the soliton $A$ and it travels at speed $4/3$ hereafter. The phase shift of the background is also $-2$.

\end{itemize}

\end{exam}

\begin{rema}
The BBS may take different boundary values at $n\to\infty$ and $n\to-\infty$ like in this example, which is not treated as a periodic BBS. 
Other irregular boundary conditions can also be considered and general solutions are calculated in the same way. 
In this section we have only dealt with the case where $L=3$. We can consider general $L$ to obtain the boundary condition in which the left side and the right side take arbitrary integer values. 
\end{rema}

\begin{verbatim}
t= 0 ::222:::::22:::::::::..................
t= 1 ::::222::::22::::::::..................
t= 2 ::::::222:::22:::::::..................
t= 3 ::::::::222::22::::::..................
t= 4 ::::::::::22::222::::..................
t= 5 :::::::::::22:::222::..................
t= 6 ::::::::::::22::::222..................
t= 7 :::::::::::::22::::131.................
t= 8 ::::::::::::::22:::.23.................
t= 9 :::::::::::::::22::..32................
t=10 ::::::::::::::::22:..131...............
t=11 :::::::::::::::::22...23...............
t=12 ::::::::::::::::::3....32..............
t=13 :::::::::::::::::.31...131.............
t=14 :::::::::::::::::.22....23.............
t=15 :::::::::::::::::.13.....32............
t=16 :::::::::::::::::..31....131...........
t=17 :::::::::::::::::..22.....23...........
t=18 :::::::::::::::::..13......32..........
t=19 :::::::::::::::::...31.....131.........
t=20 :::::::::::::::::...22......23.........
\end{verbatim}
The time evolution of 2-solitons on an irregular background is shown above. Here the frame itself moves along with the background state so that the discontinuity in the background solution seems fixed. Note that the background state itself is moving to the right at speed $1$. The symbol `:' indicates a background `\textit{1}' and `.' a zero respectively. We see that the phase shift of the background after colliding each soliton is $-2$.
\bigskip

\section{Concluding Remarks}

We first discussed the conditions under which the periodic BBSC with general box and carrier capacities is well-defined. We extended the construction of the conserved quantities to general BBSC.
Conserved quantities are useful in distinguishing background solutions from positive soliton solutions.
We then showed that the Jacobi theta function is a solution of the gauge transformed discrete KdV equation, and that the ultradiscretization of this solution corresponds to the background solution of the BBS. The author wishes to extend this method to more general functions in order to deal with negative solitons.
Finally, the BBS with irregular boundary conditions has been constructed. To obtain the solution of the system with general $L$ is a future problem. 

\section{Acknowledgements}
The author has greatly benefited from detailed discussions with Professors Jun Mada and Tetsuji Tokihiro on the conserved quantities of the BBSC.
The author also wishes to thank Professor Ralph Willox for insightful discussions on background solutions.


\begin{thebibliography}{99}

\bibitem{TOKIHIRO}
T. Tokihiro, D. Takahashi, J. Matsukidaira, J. Satsuma:
``From soliton equations to integrable systems through limiting procedures'',
\textit{Phys. Rev. Lett.} {\bf 76}(1996), 3247-3250

\bibitem{HIROTA2}
S. Tsujimoto, R. Hirota:
``Ultradiscrete KdV Equation'',
\textit{J. Phys. Soc. Jpn.} {\bf 67}(1998), 1809-1810

\bibitem{SATSUMA}
D. Takahashi, J. Satsuma:
``A soliton cellular automaton'',
\textit{J. Phys. Soc. Jpn.} {\bf 59}(1990), 3514-3519

\bibitem{TAKAHASHI2}
D. Takahashi:
``On some soliton systems defined by using boxes and balls'',
\textit{Proc. Int. Symp. on Nonlinear Theory and its Applications} (NOLTA'93)(1993), 555-558

\bibitem{TAKAHASHI}
D. Takahashi, J. Matsukidaira:
``Box and ball system with a carrier and ultradiscrete modified KdV equation'',
\textit{J. Phys. A: Math. Gen.} {\bf 30}(1997), L733-L739

\bibitem{HIROTA1}
R. Hirota:
``New Solutions to the Ultradiscrete Soliton Equations'',\\
\textit{Stud. Appl. Math.} {\bf 122}(2009), 361-376

\bibitem{WILLOX}
R. Willox, Y. Nakata, J. Satsuma, A. Ramani, B. Grammaticos:
``Solving the ultradiscrete KdV equation'',
\textit{J. Phys. A: Math. Theor.} {\bf 43}(2010), 482003, 7pp

\bibitem{kanki}
M. Kanki, J. Mada, T. Tokihiro:
``Conserved quantities and generalized solutions of the ultradiscrete KdV equation'', \textit{J. Phys. A: Math. Theor.} {\bf 44}(2011), 145202, 14pp

\bibitem{YURA}
F. Yura, T. Tokihiro:
``On a periodic soliton cellular automaton'',
\textit{J. Phys. A: Math. Gen.} {\bf 35}(2002), 3787-3801

\bibitem{MADA3}
J. Mada, M. Idzumi, T. Tokihiro:
``The exact correspondence between conserved quantities of a periodic box-ball system and string solutions of the Bethe ansatz equations'',
\textit{J. Math. Phys.} {\bf 47}(2006), 053507, 18pp

\bibitem{nagai}
This transformation was independently found by C.Gilson, A.Nagai and J.Nimmo.
(\textit{Conference talk})(December 2010)

\bibitem{IWAO}
S. Iwao, T. Tokihiro:
``Ultradiscretization of the solution of periodic Toda equation'',
\textit{J. Phys. A: Math. Theor.} {\bf 40}(2007), 12987-13021

\bibitem{TAKAHIRO}
D. Takahashi, R. Hirota:
``Ultradiscrete soliton solution of permanent type'',
\textit{J. Phys. Soc. Jpn.} {\bf 76}(2007), 104007, 6pp

\end{thebibliography}
\end{document}